# QoS Prediction for 5G Connected and Automated Driving

A. Kousaridas, R. P. Manjunath, J. Perdomo, C. Zhou, E. Zielinski, S. Schmitz and A. Pfadler

*Abstract*— 5G communication system can support the demanding quality-of-service (QoS) requirements of many advanced vehicle-to-everything (V2X) use cases. However, the safe and efficient driving, especially of automated vehicles, may be affected by sudden changes of the provided QoS. For that reason, the prediction of the QoS changes and the early notification of these predicted changes to the vehicles have been recently enabled by 5G communication systems. This solution enables the vehicles to avoid or mitigate the effect of sudden QoS changes at the application level. This article describes how QoS prediction could be generated by a 5G communication system and delivered to a V2X application. The tele-operated driving use case is used as an example to analyze the feasibility of a QoS prediction scheme. Useful recommendations for the development of a QoS prediction solution are provided, while open research topics are identified.

## I. INTRODUCTION

It is foreseen that vehicles will become more automated and wirelessly connected in the future. Cooperative perception or cooperative sensing via the wireless communication of vehicles provides a good complement to the on-board sensors by extending vision and detection ranges even when visual line-of-sight is not available. In addition, connectivity is key for cooperative maneuvers among automated vehicles to coordinate their trajectories in a safe and fast manner.

3GPP [1] and 5GAA [2] analyze various vehicle-to-everything (V2X) use cases with different performance requirements. Safety and automated driving use cases have the most demanding quality-of-service (QoS) requirements on the wireless communication system. In order to meet the demanding requirements 3GPP enhanced its 5G communication standards in Release 16 and developed a new cellular V2X standard based on the 5G NR air interface. Cellular 5G communication systems support two operation modes for V2X communication, namely V2X direct communication over the sidelink, i.e., PC5 interface (vehicle-to-vehicle (V2V)), and V2X communication over the Uu interface (vehicle-to-network (V2N)). 5G NR V2X introduces advanced functionalities to support V2X use cases with stringent QoS requirements, which could not be supported by the LTE V2X (i.e., Release 14 and 15).

In vehicular communications, the experienced QoS is affected by various factors, such as UE density, interference, mobility, handover, and roaming transitions. The avoidance of a sudden session interruption due to QoS degradation is a key requirement specifically for critical V2X services (e.g., safety, automated driving). On the other side, a feature that many V2X services have is that they can operate with different configurations, which have different QoS requirements (e.g., speed or/and video configuration of a tele-operated vehicle as analysed in section III.A). Due to this feature, the applications can continue to operate by switching to another configuration that corresponds to an alternative QoS profile (e.g., with lower QoS). The latter is provided by the communication system when the initial QoS profile cannot be provided anymore. A QoS profile is a set of QoS parameters (e.g., latency, data rate) and each QoS profile supports a specific value for each QoS parameter, as presented in [3].

These V2X service characteristics and the specific automotive requirements created the need to predict the change of the QoS level of one or more QoS parameters of an established session, as well as to provide early notifications to the vehicles about the predicted decrease or increase of the QoS. This notification can support the fast adaptation process of the V2X application.

In Release 16, 3GPP introduced a solution that allows a cellular 5G communication system to notify a V2X application of an expected or estimated change of QoS before it actually occurs [4]. This procedure is referred to as QoS sustainability analytics in 3GPP standards and helps the V2X application to decide in a proactive and safe manner if there is need for an application change (e.g., safely stop a service, adapt mode of operation of an application) when the QoS degrades. It should be mentioned that alternative QoS profile (AQP) is another complementary feature introduced in Rel. 16 that allows an application to inform the network about the list of alternative service requirements that it could operate with. This helps to avoid session interruption due to QoS degradation, since the network can quickly downgrade to another QoS. However, this feature does not provide enough time to the application for a smooth and safe adaptation, since the notification of a QoS change is sent when the actual change occurs.

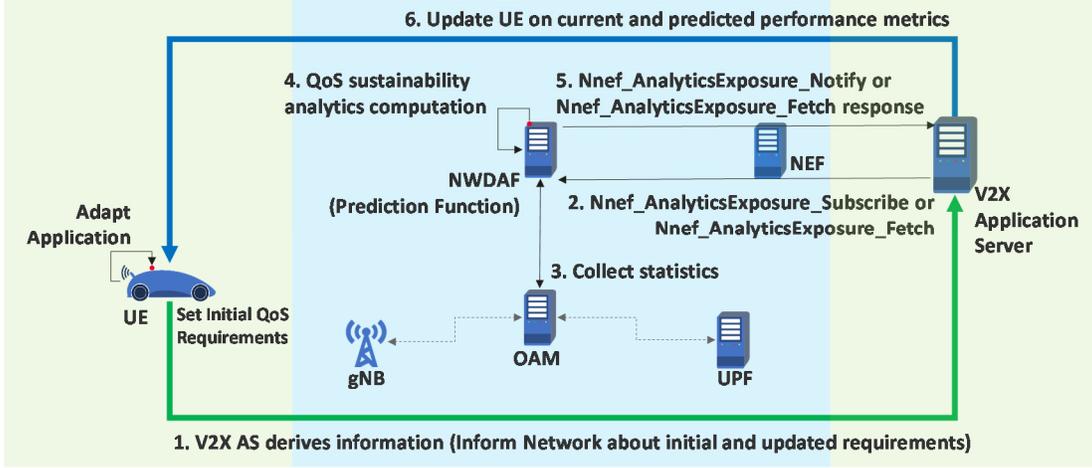

Fig. 1. Notification on QoS sustainability analytics to the V2X application server (based on 3GPP TS 23.288 [4] and 3GPP TS 23.287 [5])

This article firstly highlights the importance of QoS prediction, especially for safety-related V2X services. An overview of how real time QoS prediction of a service flow could be generated in a 5G communication system is provided. Tele-operated driving (ToD) is used as an example to study the feasibility, the performance, and different scenarios for QoS prediction determination. Useful recommendations are provided, together with an analysis of open issues.

## II. Agile QoS Adaptation

### A. Overview

Some of the future advanced V2X services, such as ToD, demand requirements on the link quality which are far beyond the conventional-consumer handheld requirements. Non-mission critical and non-safety-relevant V2X use cases (e.g., optimal route selection, navigation) can be built on best-effort connectivity schemes. However, especially mission-critical and safety-relevant types of V2V and V2N use cases (e.g., platooning, collision avoidance, ToD) demand for additional and/or specific QoS requirements referring to a specific service quality, and availability of this quality over a period of time for the V2X communication to function properly and safely.

To enable V2X services with strict QoS requirements, future cellular networks should offer an enhanced QoS-based "application adjustment assistance" mechanism and the corresponding interface, which allow tighter adjustments between the application layer and the communication system. This mechanism falls within the framework of agile QoS adaptation (AQoSA) as it allows the application to modify its configuration (e.g., move from automated assisted driving to manual mode, increase inter-vehicle gap, decrease speed), according to the QoS that can be delivered.

For each application-level configuration, a different QoS level or QoS profile may be required. V2X application can be timely notified of expected or estimated change of QoS before the actual change occurs. However, the spatiotemporal dynamics of wireless networks and high mobility of vehicles may lead to sudden QoS changes. Harsh application adjustment, especially due to a QoS degradation, e.g., sudden release of a data bearer, may affect the V2X services' performance in an inappropriate fashion, such as service discontinuity and traffic inefficiency.

For that reason, AQoSA requires an in-advance QoS notification mechanism, to enable the application for appropriate configuration adaption to the expected QoS. This notification mechanism, relies on a proper prediction of the expected future change of the provided QoS. 3GPP has recently introduced an initial mechanism that is described in section II.B. With this prediction, sudden QoS changes at the application level can be avoided or mitigated by informing the vehicles about the connectivity parameters and imminent changes to maintain safety and efficiency of V2X services.

The prediction of QoS is a forecast that includes the expected values of a QoS profile parameter and their related variances, with each parameter being dependent on the forecast duration. Based on the predicted communication QoS, V2X safety applications are able to adapt themselves to the current conditions or even to provide updated requirements.

### B. QoS Prediction in 5G Networks

3GPP has introduced a solution for the in-advance notifications on potential QoS change of the Uu interface [4]. The utilization of this procedure for V2X applications is discussed in [5].

5G communication systems provide notifications of predicted QoS changes upon request from a V2X application server (AS). The V2X AS can either subscribe to notifications or request a single notification by the network. The AS provides information to the network about a) location, b) time window to which the notification of the potential QoS change applies and c) threshold(s), indicating level(s) which, if crossed, trigger the notification that the potential QoS change can happen.

The standardized procedure to provide early QoS notifications (or "QoS sustainability" analytics) is illustrated in Fig. 1. The V2X AS requests or subscribes for analytics information on QoS sustainability provided by the network data analytics function (NWDAF) via the Network Exposure Function (NEF). NWDAF is responsible for provision of analytics and predictions. Thereinafter, the NWDAF collects statistics provided by the operations, administration, and maintenance (OAM) entity that is responsible for management plane activities, including network performance monitoring. The NWDAF verifies whether the triggering conditions are met and derives the requested analytics or prediction about expected change of QoS. QoS sustainability analytics can be provided for an indicated geographic area and time interval. The NWDAF provides response or notification on QoS sustainability to the V2X AS via the NEF. Based on the received notification by the network, V2X application adjustment may take place.

## III. QOS PREDICTION FOR TOD

### A. Tele-operated Driving and QoS Requirements

ToD has been selected to analyze the implementation and the benefits of QoS prediction, since it is an application that can operate with different configurations and QoS levels. ToD enables a remote driver to control a vehicle. The environmental conditions as observed by the vehicle sensors are transferred to the remote driver as perception data.

To realize ToD, data is exchanged through a cellular network. The quality of the transmission of perception data in uplink (UL) and control data in downlink (DL) has a tremendous impact on the quality of application. The most important QoS parameters for teleoperation are the DL latency and UL data rate.

The appropriate configuration of the perception is essential for enabling the remote control of the vehicle at the command center (CC). The configuration of the perception depends on the vehicle velocity, its environment and the QoS levels. Depending on the situation, the CC requires up to a 360° view, 8 cameras and high frame rates of up to 30 fps. A low bandwidth approach for the UL, which enables ToD in areas with low QoS levels, is the so-called Slim Uplink [6], where the vehicle transmits object data, combined with single frames. The distinct configurations with the corresponding QoS requirements in UL is provided below [6]:

- Full video upstream (360°, 8 cameras, 30 fps): 32 Mbps data rate and 40 ms latency
- Limited video upstream (360°, 5 cameras, 30 fps): 20 Mbps data rate and 40 ms latency
- Reduced video upstream (<=360°, 5-3 cameras and 30-10 fps): 3-20 Mbps data rate and 40 ms latency
- Slim Uplink (object data and single frames): 1 Mbps data rate and 40 ms latency.

In the DL, one distinguishes between direct control and indirect control. In the case of direct control, the remote driver directly controls the vehicle actuators. Indirect control refers to the control of the vehicle through the input of information such as trajectories, waypoint or high level commands. The requirements of the different DL configurations are [6]:

- Indirect control: 500 Kbps data rate and 80 ms latency.
- Direct control: 500 Kbps data rate and 40 ms latency. It has stricter latency requirements, than indirect control because the CC directly controls the vehicle actuators.

### B. QoS Prediction Horizon Determination

Adapting the ToD application to predicted QoS changes includes the adaptation of the speed or the perception configuration or the remote control settings. The selected application adaptation depends on the environment of the vehicle, the requirements of the CC, and QoS change level. The horizon of a QoS prediction is an important parameter for the application as well as for the actual QoS prediction scheme, and it is related with the application type and the adaptation action.

For instance, the speed adaptation requires a certain QoS prediction horizon ($t_{pred}$). Performing a safe stop maneuver in case of communication loss needs to be possible at all time. Emergency stop maneuvers are not desired as the user experience suffers, e.g., deceleration of 10 m/s^2. Fig. 2 depicts possible decelerations for a safe stop maneuver depending on the prediction horizon with predetermined maximum velocities. Desired decelerations are lower than 4 m/s^2.

Fig. 2 shows the minimum length of the QoS prediction horizon ($t_{pred, min}$) depending on the vehicle velocity. The higher the velocity the longer the prediction horizon has to be. Low decelerations are favored even in the case of QoS degradations which do not require a safe stop but a smooth speed adaptation. In addition to the user experience, the maneuver efficiency plays an important role as the vehicle controller can plan its maneuver in a more efficient manner when the prediction horizon is larger, in turn avoiding frequent fuel-inefficient accelerations and decelerations.

### C. QoS Prediction Technique

A QoS prediction scheme is presented for the estimation of the expected QoS of a specific service flow for a requested prediction horizon. The proposed scheme is placed at the NWDAF and enables real-time QoS prediction calculation. This provides more precise QoS prediction information, which is important for safety-related applications such as ToD. Hence, we extend the current functionality of the QoS Sustainability service (section II.B), which is based mainly on historic information. The scheme consists of two phases (Fig. 3):

- Training (Offline phase): In this phase, data are collected using network and application monitoring tools, which are processed to train and periodically update a prediction model. OAM and RRC measurements are examples sources for data collection. Different data and input features (e.g., location, cell load) need to be used for an accurate prediction of different QoS parameters. The goal of this phase is to use machine learning tools and identify the relationship of input features and their corresponding predicted parameter (label) in order to accurately estimate the expected value of a QoS parameter. This is an offline phase that is not repeated for each individual prediction request, but can be periodically repeated, according to the updates in collected data. Separate prediction models may be needed for each QoS parameter and/or for UL and DL. The trained models are provided to the Inference engine.

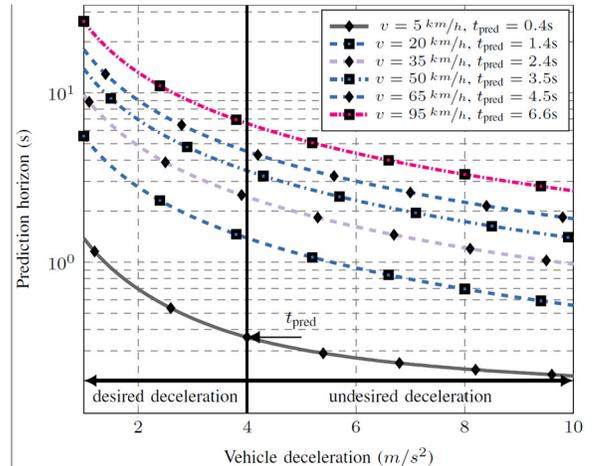

Fig. 2. ToD prediction horizon

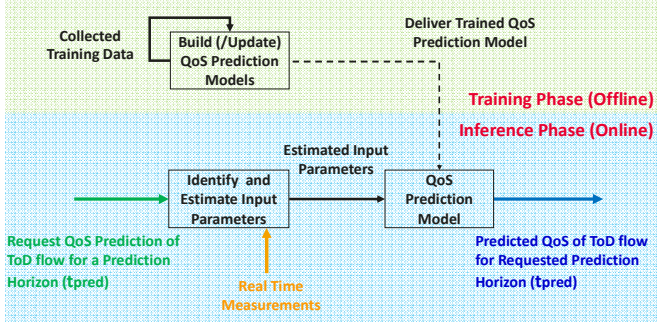

Fig. 3. Overview of QoS prediction scheme for a ToD service flow

- Inference (Online phase): This phase is activated with each prediction request. The latter, as described in section II.A includes the QoS parameter(s) that needs to be predicted, the required prediction horizon, location information etc. Based on the request type, the input features that should be used are identified and their future values are estimated for the prediction horizon that has been set by the application. Real time measurements, also for the specific service flow, are collected by the corresponding network entities (e.g., OAM) for the estimation of current and future values of the input features. At the second step of this phase, the estimated values of the input features are provided as an input to the trained QoS prediction model. The latter generates the predicted QoS of a service flow for the requested prediction horizon.

The described scheme is generic and could be used to estimate relevant QoS parameters of various applications.

## IV. EVALUATION AND DISCUSSION

In this section, we evaluate the performance of the QoS prediction technique presented in the previous section. The analysis focuses on the prediction of the UL throughput of the ToD flow, which is an important metric for high quality video streaming. The evaluation helps us to investigate the feasibility of the real time QoS prediction at service flow level, to derive useful observations, and to identify open questions that are summarized in Section V.

### A. Simulation Scenario

A system-level simulation setup for the ToD use case is used for data collection to train the QoS prediction model and evaluate its performance. The network and mobility simulations are carried out using the ns-3.33 network simulator [7] and the vehicular mobility simulator SUMO [8], respectively. The 3GPP evaluation methodology defined in [9] is used for the simulation environment.

Specifically, an urban grid road configuration is used (Annex A and Table 6.1.1-1 in [9]) and the network consists of three macro sites, using tri-sector antenna. The macro base stations (BSs) have inter site distance (ISD) of 500 meters and they operate at the 2160 MHz center frequency with a 20 MHz system bandwidth, employing frequency division duplex (FDD). Type 2 (passenger) vehicles [9] are used with a single isotropic antenna that have an average speed of 50 kmph. Large-scale signal propagation is modelled using the building-aware path loss model ITU-R P.1411 for line-of-sight and for non-line-of-sight [10].

In the simulation, a ToD vehicle runs a ToD application and interacts with a ToD application server (CC), via the Uu interface. The ToD application UL data rate is 20 Mbps for the video streaming. Several non-ToD (NToD) vehicles send UL data packets to a NToD application server, using an aperiodic traffic models (inter-packet arrival time: an exponential random variable with the mean of one second, packet size: 1012 bytes). The core network is comprised by a packet gateway, which connects the BSs to the NToD and ToD application servers.

The BSs allocate resources using an independent round robin scheduler per sector and a best effort QoS bearer for both ToD and NToD traffic, i.e., no service differentiation at the radio access network (RAN). It should be noted that in a real environment the ToD service needs a guaranteed bearer. However, in our analysis we have selected to treat all types of traffic with a best effort QoS bearer to create more volatile traffic conditions for the ToD service and thus more variance in the collected data for our QoS prediction analysis.

To allow data collection for varying network loads, several simulation runs are performed with different amount of NToD vehicles (i.e., 0, 5, 15, 30, 40, 50, 70, 80, 100, 130 and 160 NToD vehicles) and different mobility traces of the vehicles. The ToD UL throughput is affected by the experienced channel, UL inter-cell interference and scheduling. The effect of different background traffic levels on the ToD UL throughput are:

- No background traffic: the ToD vehicle maintains a constant 20 Mbps throughput.
- Low background traffic: the ToD vehicle maintains a 20 Mbps throughput with sporadic variations down to 15 Mbps.
- Medium background traffic: the ToD UL throughput varies between 20 and 10 Mbps.
- High background traffic: the ToD UL throughput varies between 20 Mbps down to 5 Mbps.

### B. QoS Prediction Performance

#### 1) Prediction Technique Configuration

For the prediction of the ToD UL throughput, the random forest (RF) prediction algorithm for regression [11] is used to realise the scheme presented in Section III.C. The RF is a versatile model that is resistant to overfitting issues, different machine learning algorithms could be used such as kernel ridge regression or neural network based predictors.

TABLE I
ToD VEHICLE UL THROUGHPUT - PREDICTION ACCURACY OF TRAINING
(OFFLINE PHASE)

| Prediction Configuration | Selected Input Features | Output: UL Throughput | | |
|---|---|---|---|---|
| | | Mean Abs. Error (kbps) | Std. dev. Abs. Error (kbps) | Mean Abs. Percentage Error |
| ToD service and network related features | | | | |
| T1 | All Features | 45 | 241 | 0.005 |
| T2 (ToD Cell features) | V_C1, D_C1, Pos_X, Pos_Y, ToD_Dist | 95 | 452 | 0.010 |
| ToD service related features | | | | |
| T3 | Pos_X, Pos_Y, ToD_Dist | 1528 | 2053 | 0.140 |
| Network related features | | | | |
| T4 | V_C1, V_C2, V_C3, D_C1, D_C2, D_C3, NToD_Dist | 607 | 1637 | 0.070 |
| T5 | V_C1, V_C2, V_C3, D_C1, D_C2, D_C3 | 785 | 1873 | 0.093 |
| T6 (ToD Cell features) | V_C1, D_C1 | 1506 | 2581 | 0.220 |

The dataset collected from various simulation scenarios that vary in terms of NToD background traffic and mobility traces, as presented in the section IV.A, and is used for the training of the RF-based QoS prediction model. The dataset consists of 1281228 samples and the following input features:

- location of the ToD vehicle (x,y coordinates) (*Pos_X*, *Pos_Y*),
- distance of the ToD vehicle to BS (*ToD_Dist*),
- number of vehicles in each cell (*V_C1, V_C2, V_ C3*),
- data rate demand of the NToD vehicles in each cell (*D_C1, D_C2, D_C3*),
- reciprocal of the sum of distances between the NToD vehicles not attached to the ToD's serving cell and the ToD's serving cell (*NToD_Dist*).

This comprises of twenty-two features in total, considering three cells simulation scenario (ToD vehicle cell and two neighboring cells) and three sectors per site. The UL throughput (in bps) of the ToD vehicle is predicted by the RF model.

*2) Prediction Performance of Training Phase*

Firstly, we analyze the training phase of the QoS prediction scheme (Fig. 3) and specifically the prediction accuracy of the trained prediction model. The metrics to evaluate the prediction accuracy are the following: a) Mean absolute error (MAE), b) Standard deviation of the absolute error, c) Mean absolute percentage error (MAPE).

Table I shows the RF performance for the prediction of ToD UL throughput, considering randomly chosen two-thirds of the data for the training of the QoS prediction model and one third for testing. Different combinations of input features, referred as prediction configurations, are chosen to study their impact on an accurate prediction of the ToD UL throughput.

The results show that the combination of ToD service and network related features can provide relatively good prediction accuracy (configuration T1 and T2). These features are able to capture the performance of the cell to which the ToD vehicle is attached to, the performance of ToD vehicle, and the inter-cell interference from neighboring cells. Since the simulations and hence the dataset is based on using a round-robin MAC scheduler, the number of vehicles along with their demand has played a significant role in accurately predicting the UL throughput. By reducing the features to only ToD specific information, such as, ToD vehicle's location and distance information (configuration T3), it is not enough for an accurate prediction. In multi-cell communication scenarios, when the transmissions of vehicles in one cell can be influenced by transmissions in neighboring cells, metrics capturing relevant network load and possible interference information is essential. Based on prediction accuracy results with mainly cell-specific features (configuration T4, T5, T6), we notice significant accuracy deviation for all the considered accuracy metrics. Cell-specific features combined with specific ToD service information help make relatively accurate predictions.

The development of a precise prediction model, at the training phase, should include the key impacting features and should capture various network situations. This is the first required step for the derivation of an accurate QoS prediction. A good selection of input features increases the accuracy of the prediction of a QoS parameter.

*3) Performance of Inference in known scenario*

Thereinafter, we focus on predicting real time UL throughput for a certain prediction horizon (Inference phase, Fig. 3). The prediction model should be fed with estimated future values of the input features for the specific horizon. We evaluate the inference phase for two kinds of input scenarios:

- Perfect input estimates: corresponds to the case when the estimates of the input features during the prediction horizon are accurate.
- Imperfect input estimates: corresponds to the case when the estimates of the input features during the prediction horizon are inaccurate.

In this section, we test the prediction model in a known scenario. This corresponds to the case when the trained model has knowledge of the testing scenario. For example, model trained for a peak traffic scenario and tested for a similar scenario. The motivation for considering the above kinds of scenarios is to understand the performance of inference phase by comparing an ideal case (perfect input estimates) with, probably, a more realistic one with imperfect input estimates.

Fig. 4 shows an example of the inference for the prediction of the ToD UL throughput for a known scenario using both perfect and imperfect input estimates. A maximum of seven seconds prediction horizon is needed by a ToD service, according to Fig. 2. The inference is applied for one of the conducted simulations with 80 NToD vehicles, using the RF model trained with configuration T1. In the case of perfect input features, the predicted ToD UL throughput for a seven seconds prediction horizon is very close to the true UL throughput, due to the well trained and designed QoS prediction model.

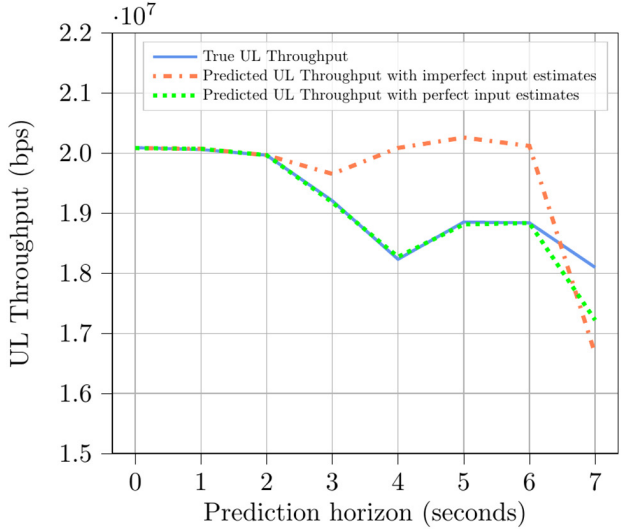

Fig. 4. Example ToD UL throughput prediction for prediction horizon ($t_{pred}$) of seven seconds – Inference in known scenario

However in reality, we need some techniques to estimate future values of the input features for the specific horizon. In many cases the spatiotemporal dynamics of traffic exhibit a roughly periodic pattern. Hence, in our example a time-series approach, based on the auto-regressive integrated moving average (ARIMA) model is used to estimate the future values of the number of vehicles and data rate demand input features, for the requested period up to seven seconds. Future location information and current velocity of the ToD vehicle can be readily available by the ToD application. As Fig. 4 presents, the predicted ToD UL throughput, with ARIMA-based (imperfect) input estimates is close to the ideal case with perfect input estimates. For the first three seconds of the prediction horizon the prediction accuracy is high. After the fourth second there is an increase of the prediction error, due to less accurate estimation of future values of the input features.

Depending on the nature of key features, their respective estimation may have different degrees of uncertainty. This is an additional reason why the key input features should be carefully selected. In general, different techniques can be used to estimate future values of the input features for a specific horizon. If these techniques are combined with relevant context information, then uncertainty can be reduced.

*4) Performance of Inference in unknown scenario*

Finally, we test the prediction model in an *unknown* scenario. This corresponds to the case when the trained model has no knowledge of the testing scenario.

Requesting the predicted ToD UL throughput for a specific prediction horizon in an unknown testing scenario is affecting the prediction accuracy. Fig. 5 presents an example of the predicted UL throughput in the case that the conducted simulation with 80 NToD vehicles is an unknown scenario. The configuration of the simulation environment and the RF algorithm is the same as in the known scenario (Fig. 4). The difference is that the 80 NToD vehicles scenario is not part of the training data used to build the RF-based ToD UL throughput prediction model. This fact increases the uncertainty for the QoS prediction, as completely new situation appears in terms of different load and UL inter-cell interference level.

Especially, after the fourth second there is an increase of the QoS prediction error, even with perfect input features. This level of accuracy may not be useful for some applications. Hence, techniques are needed to reduce the impact of unknown scenarios in QoS prediction, since they will be met in reality.

## V. CONCLUSIONS AND FUTURE CHALLENGES

The work presented in this article focuses on analyzing the components and functions required for the QoS prediction of a service flow for a requested prediction horizon, using real time measurements. ToD has been used as an example to provide useful insights and to identify open research topics. Based on the examples presented in Section IV.B, it is shown that an accurate real-time QoS prediction for a service flow is feasible. This can provide more precise and live QoS prediction information to an application and thus have safer and more efficient application adaptations to the QoS changes.

The identification of the key input features that affect the expected QoS performance is the first step needed to train an accurate QoS prediction model. For instance, for the ToD UL throughput prediction the location information of the vehicles in combination with load demand information from the serving cell as well as the neighboring cells as input features enables a high prediction accuracy (i.e., MAPE of 0.005, Table I). Neighboring cells' information (i.e., data rate demand and average distance of the UEs not attached to the BS to which the ToD vehicle is attached) is helpful to capture the impact of the interference on the UL transmissions of the ToD vehicle and thus to achieve a more accurate prediction. Additional features could be considered such as the variation of the channel conditions, however, an exhaustive analysis of these aspects is beyond the scope of this article. The acceptable prediction error levels for each application need to be investigated for precise

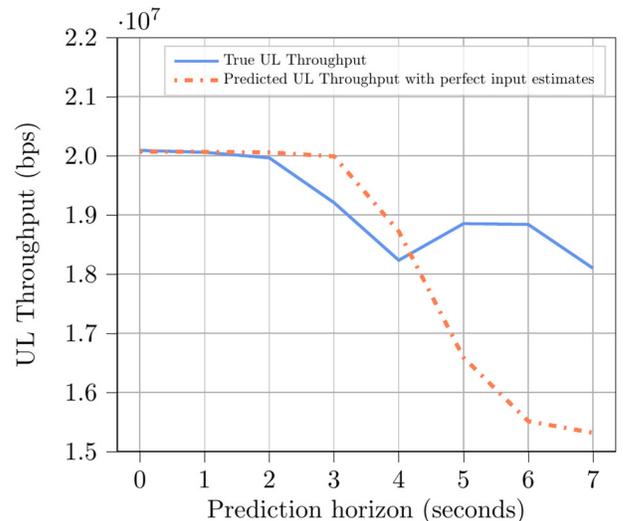

Fig. 5. Example ToD UL throughput prediction for prediction horizon ($t_{pred}$) of seven seconds – Inference in unknown scenario

evaluation of prediction performance.

Each application requires an appropriate prediction horizon and is an additional factor that affects the prediction accuracy and should therefore be carefully selected. The longer the prediction horizon, the higher the introduced uncertainty. It becomes more difficult to estimate the future values of the selected input features in the inference phase, and this uncertainty is transferred into the actual QoS prediction. Time-series may be able to capture the trend behavior and still keep acceptable error levels for short-term estimations of the input features. But further study is needed and whether the exploitation of relevant context information (e.g., road topology), in combination with a time series method, can help to reduce this uncertainty.

A prediction model can be built using training data from various scenarios at the training phase. However, it is difficult to consider all potential scenarios beforehand. Hence, the selection of algorithmic tools that behave adequately in unknown scenarios is necessary; advanced techniques, such as online learning and federated learning, which can cope with the dynamicity in the network with limited data sharing between different network entities could be investigated.

The QoS prediction of a sidelink interface and in multi-domain environments (e.g., Multi–MNO) are additional topics for future research. Sidelink QoS prediction can be challenging due to more complex scenarios compared to Uu, varying sidelink-specific spatiotemporal dynamics (e.g., high mobility, fast topology change), vehicle attributes (e.g., diverse antenna designs, low antenna heights) and simultaneous sidelink transmissions, especially in dense environments.

**Dr. Apostolos Kousaridas** received his Ph.D. from the Department of Informatics & Telecommunications at the University of Athens. He is a principal research engineer of the Huawei Research Center in Munich, contributing to the design of 5G communication systems and beyond. He has disseminated over 50 publications and he has contributed to ETSI, 3GPP, and 5GAA. He is serving as vice-chair of the 5G-PPP Automotive WG. His research interests include vehicular communications, wireless networks and artificial intelligence.

**Ramya Panthangi** received her M.Sc. degree in communications engineering from Technical University Munich, Germany. Currently, she is working toward Ph.D. degree with Huawei Munich Research Center and Technical University Berlin. Her research interests include vehicular communication networks and machine learning for vehicular communication networks.

**José Perdomo** received his M.Sc. degree in pervasive computing and communications for sustainable development from Luleå University of Technology, Sweden. Currently, he is a Ph.D. candidate with Huawei Munich Research Center and Polytechnic University of Valencia (UPV). His research interests include artificial intelligence, resource allocation and predictive quality of service.

**Dr. Chan Zhou** received the Dipl.-Ing. degree in computer science from the technical university Berlin, Germany, in 2003, and a Ph.D. degree in electrical engineering in 2009. In 2010 he joined Huawei European Research Center in Munich. His expertise spans from 5G research towards prototyping and demonstrations for various 5G use cases. He is currently the vice chair of the System Architecture and Solution Development Working Group in 5GAA.

**Ernst Zielinski** received his Dipl.-Ing. degree in electrical engineering (focusing on communication systems) from the Technical University Dortmund, Germany, in 2002. He started his career at Nokia Research Center in 2003, and in 2008 joined Research in Motion, Bochum. In 2014, he joined Volkswagen Infotainment, a subsidiary of the Volkswagen group, and has since been working in the field of automotive wireless communication systems. He also represents Volkswagen in the 3GPP and 5GAA. His expertise includes physical layer algorithms and system architectures for wireless communication systems.

**Steffen Schmitz** received his Dipl.-Ing. degree in electrical engineering (focusing on high-frequency engineering) from the University Duisburg-Essen, Germany, in 2003. Before joining Volkswagen Infotainment, a subsidiary of the Volkswagen group, he had been with Nokia and Blackberry. As a specialist for wireless connectivity he is representing Volkswagen in the 3GPP and 5GAA.

**Andreas Pfadler** received the M.Sc. degree in telecommunication engineering from the Polytechnic University of Catalonia (UPC), Barcelona. He is currently working toward the Ph.D. degree with Volkswagen AG and the Technical University of Berlin. His research interests include antenna design, predictive quality of service, new waveforms, and wave propagation.